\documentclass[sigconf]{acmart}

\renewcommand\footnotetextcopyrightpermission[1]{} 
\def\BibTeX{{\rm B\kern-.05em{\sc i\kern-.025em b}\kern-.08emT\kern-.1667em\lower.7ex\hbox{E}\kern-.125emX}}

\usepackage[linesnumbered,ruled]{algorithm2e}

\usepackage{color}
\usepackage{soul}

\begin{document}

\title{UltraShare: FPGA-based Dynamic Accelerator Sharing and Allocation}

\lhead{}
\rhead{}

\author{Siavash Rezaei}
\affiliation{%
  \institution{University of California, Irvine}
  \city{Irvine}
  \state{California}
  \country{USA}
}

\author{Eli Bozorgzadeh}
\affiliation{%
  \institution{University of California, Irvine}
  \city{Irvine}
  \state{California}
  \country{USA}
}
 
\author{Kanghee Kim}
\affiliation{%
  \institution{Soongsil University}
  \city{Seoul}
  \country{South Korea}
}
%
\begin{abstract}
Despite all the available commercial and open-source frameworks to ease deploying FPGAs in accelerating applications, the current schemes fail to support sharing multiple accelerators among various applications. There are three main features that an accelerator sharing scheme requires to support: exploiting dynamic parallelism of multiple accelerators for a single application, sharing accelerators among multiple applications, and providing a non-blocking congestion-free environment for applications to invoke the accelerators.
In this paper, we developed a scalable fully functional hardware controller, called UltraShare, with a supporting software stack that provides a dynamic accelerator sharing scheme through an accelerators grouping mechanism. UltraShare allows software applications to fully utilize FPGA accelerators in a non-blocking congestion-free environment. Our experimental results for a simple scenario of a combination of three streaming accelerators invocation show an improvement of up to 8x in throughput of the accelerators by removing accelerators idle times.

\end{abstract}



\settopmatter{printacmref=false}
\maketitle

\section{Introduction}
In the era of big data and computational/data-intensive applications, heterogeneous architectures are promising platforms to tackle high computational loads. FPGAs have been assisting CPUs as custom hardware accelerators with their fine-grain programmable hardware resources. FPGAs are considerably low-power and researchers have shown an order of magnitude acceleration for various applications \cite{Ting2018, catapult, Jason_dac18}. 

Streaming applications, such as image/video processing application, real-time vision algorithms, and network packets encryption algorithms, are in the category of FPGA-friendly data-intensive applications  \cite{Reconfig2016}. These applications process a stream of input data on FPGA accelerators and send a stream of output data.  Multi-level memory hierarchy model that is used in OpenCL-based platforms fails to meet their high demanding data throughput \cite{STAccel}. In these platforms, a global memory is shared between the host and the accelerator. While the host is sending data to (or receiving data from) the global memories, they cannot be accessed by the accelerators. Vice versa, when the accelerators are accessing data, global memories cannot be accessed by the host. This results in a considerable performance degradation, especially for streaming applications. Ruan \cite{STAccel} has shown that a point to point data transfer from main memory to FPGA BRAMs can significantly improve the performance of the streaming accelerators comparing to an OpenCL hierarchy memory model.

When multiple accelerators on FPGA are deployed to accelerate various streaming applications, shared hardware resources incur more stringent constraints on high throughput data movement between FPGA and the main memory. While point-to-point data transfer between FPGA local memory and main memory is necessary, a scalable and efficient high-throughput data movement infrastructure between the host and the FPGA accelerators is required. This paper proposes such a framework to  enable accelerator sharing among multiple streaming applications.

Accelerator allocation is a crucial task when multiple accelerators on an FPGA device are shared among various applications. 
For every single request from the host, an accelerator on the FPGA is allocated to the request until the result is sent back to the host.
To the best of our knowledge, all the current FPGA accelerator frameworks \cite{SDAccel, IntelSDK, STAccel, riffa, JetStream, fflink} follow a static accelerator allocation scheme; it means that software developers have to exactly specify the target accelerators for any access request in the software code. This can lead to a poor utilization of accelerators when being shared among various applications. 

In this paper, we propose UltraShare, an open-source RTL level framework, which enables a scalable and efficient FPGA-based accelerator sharing. Unlike the currently available frameworks, UltraShare invokes a dynamic accelerator allocation to requests. UltraShare also reduces the idle time of the accelerators through deploying an accelerator grouping mechanism which results in a considerable improvement in the performance of FPGA accelerators. In addition, the data size and throughput can vary from one group of accelerators to another. UltraShare provides a fair or priority-based data transfer to/from accelerators.  We briefly summarize the contributions of this paper as follow:

\begin{itemize}
    \item For the first time, we introduce a non-blocking FPGA-based accelerator sharing framework, called UltraShare, by proposing a hardware controller to enable dynamic sharing,
    \item we propose an accelerator grouping architecture to enable efficient access to accelerators shared among multiple streaming applications,
    \item we propose an algorithm for dynamic accelerator allocation and data transfer scheduler for streaming applications,
    \item We developed UltraShare in Verilog hardware programming language which makes it compatible with all the FPGA vendors and RTL synthesis tools. UltraShare is an open-source framework and can be used and contributed by other research groups,
    \item We evaluated UltraShare with the standard IP-cores interfacing standard AXI-Stream protocol on a Xilinx\textsuperscript{\textregistered} Virtex 7 FPGA.
\end{itemize}

The rest of this paper is organized as follows. In section \ref{Background} we introduce FPGA-based accelerator sharing and investigate the currently available frameworks. In section \ref{multi_accelerator_management} we introduce UltraShare and its composing components. Section \ref{Results} represents our experimental results to show the functionality of UltraShare and the impact of dynamic allocation and accelerator grouping in the performance of the FPGA accelerators. Finally, section \ref{Conclusions} concludes the paper.

\section{Background and Motivation}\label{Background}
Unlike GPUs that provide multiple instances of a single type accelerator, FPGAs provide a single platform that can implement multiple accelerators regardless of their types.
In data centers and edge computing platforms,  different applications use available accelerators on FPGAs to accelerate their computational intensive kernels. The accelerators must be accessed through an underlying infrastructure that interfacing host and FPGA accelerators.
There are currently a few numbers of academic \cite{riffa, JetStream, fflink, STAccel, mqmaiICCD18} and industrial \cite{SDAccel, IntelSDK} frameworks that are designed to connect a host to the FPGA accelerators. However, they either do not support or fail to provide a seamless interface to multiple accelerators accessed simultaneously by various applications.
An efficient multi-accelerator management is necessitated to first provide the possibility of accessing accelerators by the host applications and then minimizing the access overheads and accelerator idle times.

One of the most important requirements of a multi-accelerator system is the capability of {\em sharing} accelerators among different host applications.
An efficient accelerator sharing mechanism addresses the following features:

\begin{itemize}

\item \textbf{Exploit dynamic parallelism:}
All the requests from one application are distributed among the available accelerators. Thus, through a dynamic request to accelerator allocation, an application can benefit from all the available accelerators to reach the maximum possible performance.

\item \textbf{Sharing accelerators among multiple applications:}
Multiple application can share a single accelerator. Regardless of the source of submitting a request, accelerators must be exploited by the requests.

\item \textbf{Non-blocking congestion-free accelerators:}
While an acceleration request is getting processed, other applications are able to submit requests for the same accelerator and would not be blocked until the accelerator is idle.

\end{itemize}

Through a static accelerator allocation \cite{riffa, JetStream, fflink, STAccel, SDAccel, IntelSDK} exploiting parallelism is not practically possible, since applications are not aware of the status of all the accelerators on FPGAs. 


While using software level blocking mechanisms, like semaphore, allows some of the currently available frameworks support sharing accelerators among multiple threads in one application \cite{riffa, JetStream, fflink, STAccel}, they fail to support accelerator sharing among multiple applications. OpenCL based frameworks like Xilinx\textsuperscript{\textregistered} SDAccel \cite{SDAccel} and Intel SDK \cite{IntelSDK}, support sharing accelerators among multiple applications.
However, all the OpenCL-based platforms do not support a non-blocking congestion-free accelerator invocation. In the OpenCL programming model the flow of invoking the accelerators is constructed of three main routines: 1) writing input data to the FPGA board, 2) initiating the accelerator, and 3) reading back the results from FPGA board to the host. When the first routine is called the corresponding accelerator will be reserved for that process and no other application can call the accelerator until the third routine is completed. In addition, all the existing frameworks rely on user accelerator allocation and lack a hardware controller to enable dynamic acceleration allocation and sharing on FPGA devices. 

\begin{figure}
\centerline{\includegraphics[scale=0.42]{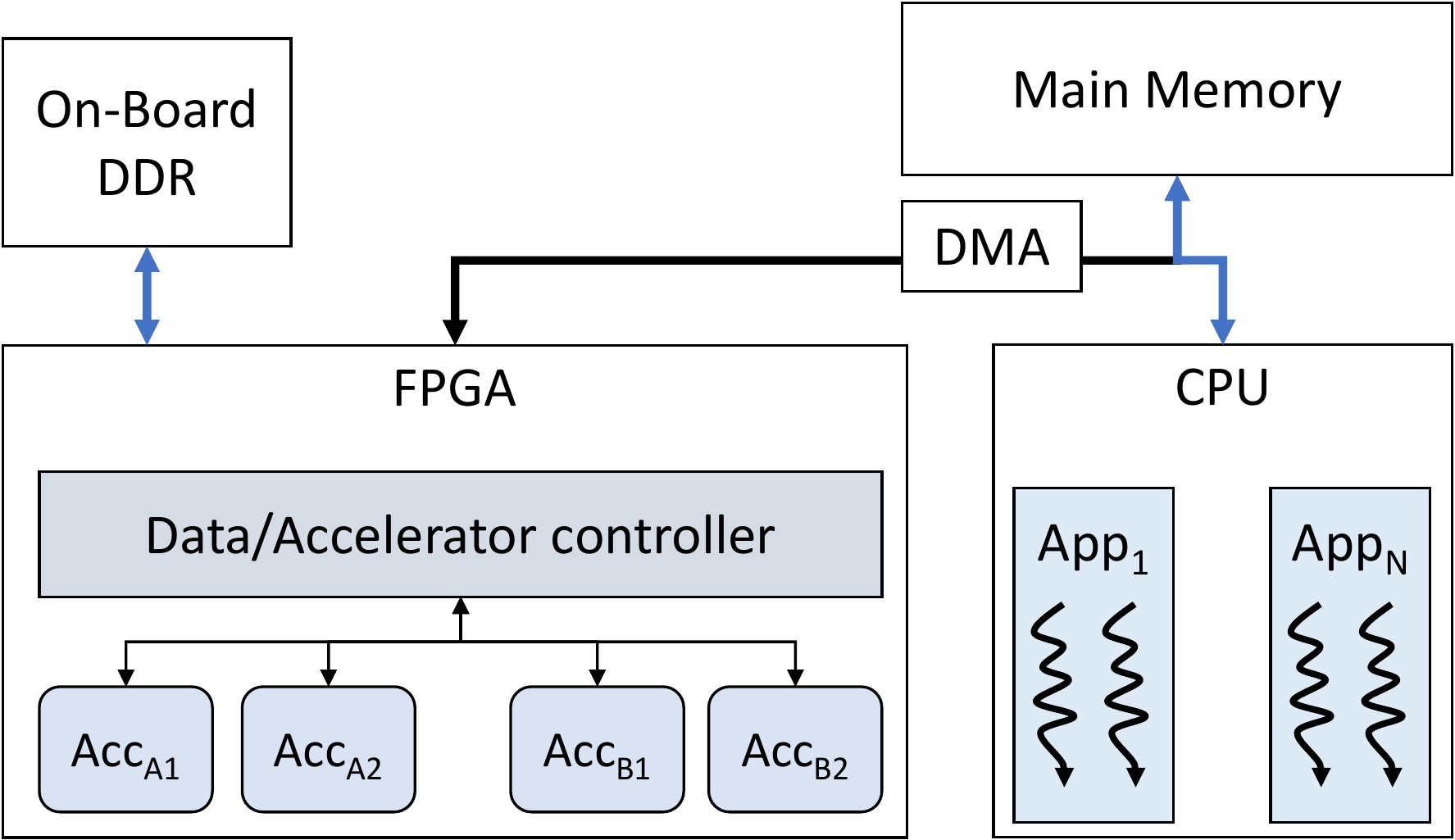}}
\caption{Target platform}
\label{platform}
\end{figure}

In this paper, we focus on FPGA-based multi-accelerator systems for streaming applications. Our target platform, as shown in Figure \ref{platform}, is composed of a host with the main memory which has access to multiple accelerators on an FPGA board through a PCIe connection.  We consider a single-level buffer-based data movement from the main memory to FPGA Block RAMs (BRAMs). In our target platform, multiple accelerators of various types or functions are deployed on the FPGA and there can be multiple instances of the same type of accelerators on FPGA device. This is referred to as a group of accelerators of the same type. 

In this paper, we propose an adaptable scalable hardware controller, called UltraShare, to address the three aforementioned features of accelerator sharing along with an efficient data movement between the host and FPGA.
In most of the available frameworks, a finite-state machine in the software stack keeps the host interacting with accelerators and does not allow a new request to be issued for the same accelerator. To address a congestion-free non-blocking accelerator sharing we employ a single command-based mechanism for each accelerator request that eliminates all the interactions between a host and accelerators after the request is issued.
This mechanism is used in NVMe \cite{NVMe} protocol and adapted by MQMAI \cite{mqmaiICCD18} for FPGAs to minimize the interaction between a host and FPGAs.
The command-based mechanism allows a single command to contain all the information needed by the accelerators to accomplish an FPGA-acceleration.
Thus, UltraShare hardware controller receives a stream of commands from the host. UltraShare dynamically allocates the commands to the accelerators. UltraShare is fully hardware-based and is located between accelerators and PCIe data interface.

\section{Multi-accelerator Management}\label{multi_accelerator_management}

\begin{figure*}
\centerline{\includegraphics[scale=0.7]{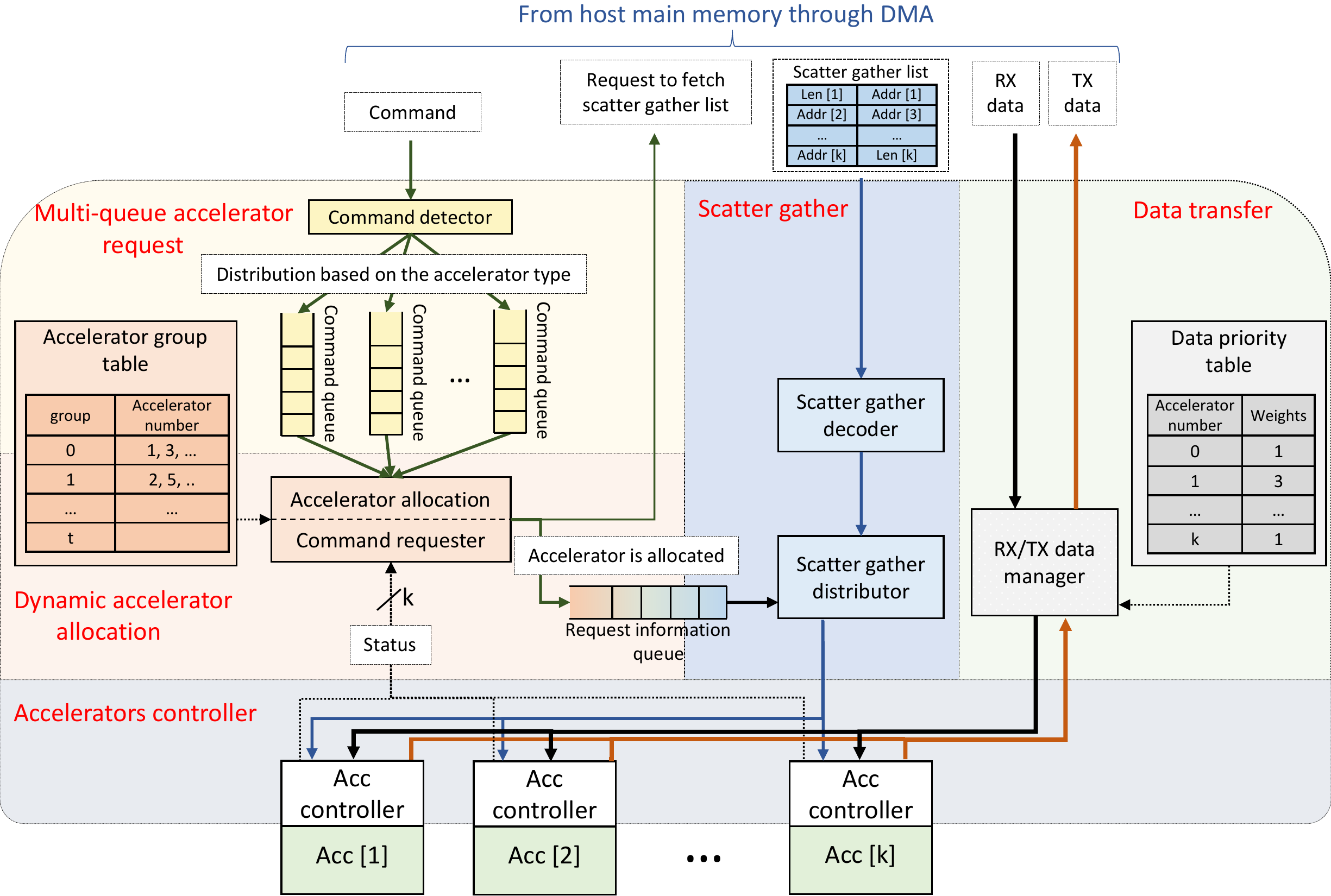}}
\caption{UltraShare hardware parts and components}
\label{hardware_arch}
\end{figure*}

UltraShare is composed of five main parts (Figure \ref{hardware_arch}): 1) multi-queue accelerator request, 2) dynamic accelerator allocation, and 3) scatter-gather, 4) accelerators controller, and 5) data transfer. The inputs to the hardware controller are streams of commands from the host main memory. UltraShare is an interaction-free framework that after receiving commands from the host all the operations are executed and initiated in the hardware controller. 
In the following we explain different components of each part in detail. 

\subsection{Multiple Command Queues}\label{single_vs_multi_queue}

Accelerator allocation is responsible for assigning commands to accelerators dynamically.
In a single-queue non-grouping mechanism, always the command at the head will be processed. If there is no accelerator available for this command, it would block the rest of commands to be processed. Thus, the single-queue mechanism may result in a severe accelerator underutilization due to the blocking requests among multiple applications requesting accelerators of different types.
In order to tackle this, UltraShare proposes an accelerator allocation based on an accelerator grouping mechanism instead of a single-queue non-grouping scheme.
Each group of accelerators has a unique command queue FIFO. Accelerator allocation part includes the following components.

\textbf{Command Detector:}
A command includes all the information required to process the associated request without any interaction with the host. This information includes: 1) command ID, 2) CPU core ID that submitted the request, 3) requested accelerator type, and 4) all the addresses and lengths of scatter-gather lists for all the inputs and outputs.
When a command arrives, the command detector pushes the command into one of the command queues.
To determine the target command queue, the command detector uses the command type field and a software re-configurable accelerator grouping table. In this paper, we only consider a one-level accelerator grouping which is based on the accelerators types, however, UltraShare framework allows more sophisticated strategies, e.g. a two-level priority-based grouping the first level of which is based on the priority of the accelerators and the second level is based on the accelerators types. In this regard, some of the accelerators can be reserved for high-priority requests.

\textbf{Command Queues:}
Command queues are simple FIFOs implemented with BRAMs. For each group of accelerators, there is one dedicated command queue.

\subsection{Dynamic Accelerator Allocation}

\textbf{Accelerator Allocation Unit:}
The main unit of the dynamic accelerator allocation part is an accelerator allocator unit. This unit assigns an accelerator to the command which is on the head of a command queue. The accelerator allocator travels between the queues in a round-robin scheduling mechanism. If there is no accelerator available for a selected command, the next command queue will be selected. In an accelerator type-based grouping mechanism, if an accelerator is idle and there is at least one command available for that type of accelerator, the command will be assigned to the accelerator.

Algorithm \ref{acc_allocation} represents the pseudo-code of the accelerator allocation unit.
The inputs to the accelerator allocation unit are: 1) the status of all accelerators, and 2) the output of \textit{accelerator group table} that represents the mapping of accelerator numbers to the accelerator groups.
Accelerator group table is re-configurable through configuration commands from the host applications. It allows regrouping of accelerators without the FPGA re-configuration cost.

In Algorithm \ref{acc_allocation}, $k$ is the number of accelerators and $t$ is number of accelerator groups. $acc\_status$ is an array with the size of the number of accelerators. If $acc\_status[i]$ is equal to $1$ then accelerator $i$ is idle and can be allocated to a request. $acc\_map$ is a two-dimensional matrix. Each accelerator group is related to a row of $acc\_map$ matrix and each accelerator is related to a column. If an element of a row is $1$, then the accelerator number with the column number is in that accelerator group. In Algorithm \ref{acc_allocation}, accelerator allocation unit travels in a round-robin scheduling among the command queues to allocate accelerators. For each selected queue ($Q$) the array of idle accelerators are determined. If this array has more than one $1$, the command on the head of the selected queue will be assigned to one of the idle accelerators. We always pick the idle accelerator with the smallest accelerator number first (rightmost position in the $idle\_acc$ array).
\begin{algorithm}
\caption{Accelerator allocation}
\label{acc_allocation}
\KwIn{bool acc\_status[k],  bool acc\_map[t][k]}
\KwOut{bool allocated\_acc[k]}
    $Q \gets 0$\;
    \While{true}{
    select queue number $Q$\;
    $idle\_acc \gets acc\_status\ \&\ acc\_map[Q]$\;
    \If{$idle\_acc \neq 0$}{
      In $idle\_acc$ keep the rightmost \textbf{1} and make the rest of the bits \textbf{0}\;
      $allocated\_acc \gets idle\_acc$\;
    }
    $Q \gets next\ Q$\;
    }
\end{algorithm}

\textbf{Accelerator Group Table:}
Accelerator group table is a lookup table that provides the information of matching accelerators to the accelerator groups. This lookup table is re-configurable through software APIs, and a user can regroup accelerators, remove an accelerator from a group, or add accelerators to different groups.

\textbf{Command Requester:}
After allocating an accelerator to a command, \textit{command requester unit} submits a request to the DMA to fetch input data (RX) and output data (TX) scatter-gather lists. When the DMA request is submitted, a signal will be sent to the \textit{accelerator allocation unit} to process the next command.

\textbf{Request Information Queue:}
The information related to each processed command (including the allocated accelerator number, length of the scatter-gathers list) will be stored in a queue to be used when the associated scatter-gather lists arrive.

\subsection{Scatter-Gather}
The scatter-gather part is responsible for receiving and decoding scatter-gather lists, and distributing them into their associate accelerators. The data management part is composed of the following components.

\textbf{Scatter-Gather Decoder:}
A scatter-gather list is constructed of a list of memory page addresses with their associated data length; while usually, the length of the first and last scatter-gather is less than a memory page size, the length of the other scatter-gathers is equal to the page size. To shorten the size of the scatter-gather list, we compact it with skipping the length of middle scatter-gathers. When a scatter-gather list arrives, \textit{scatter-gather decoder} extracts pairs of addresses and lengths from a scatter-gather list. We call each pair of address and length a scatter-gather element.

\textbf{Scatter-Gather Distributor:}
The inputs to \textit{scatter-gather distributor} are scatter-gather elements and the information from \textit{request information queue}. Using the information in the request information queue, scatter-gather elements are submitted to the allocated accelerator controller.

\begin{figure*}
\centerline{\includegraphics[scale=0.67]{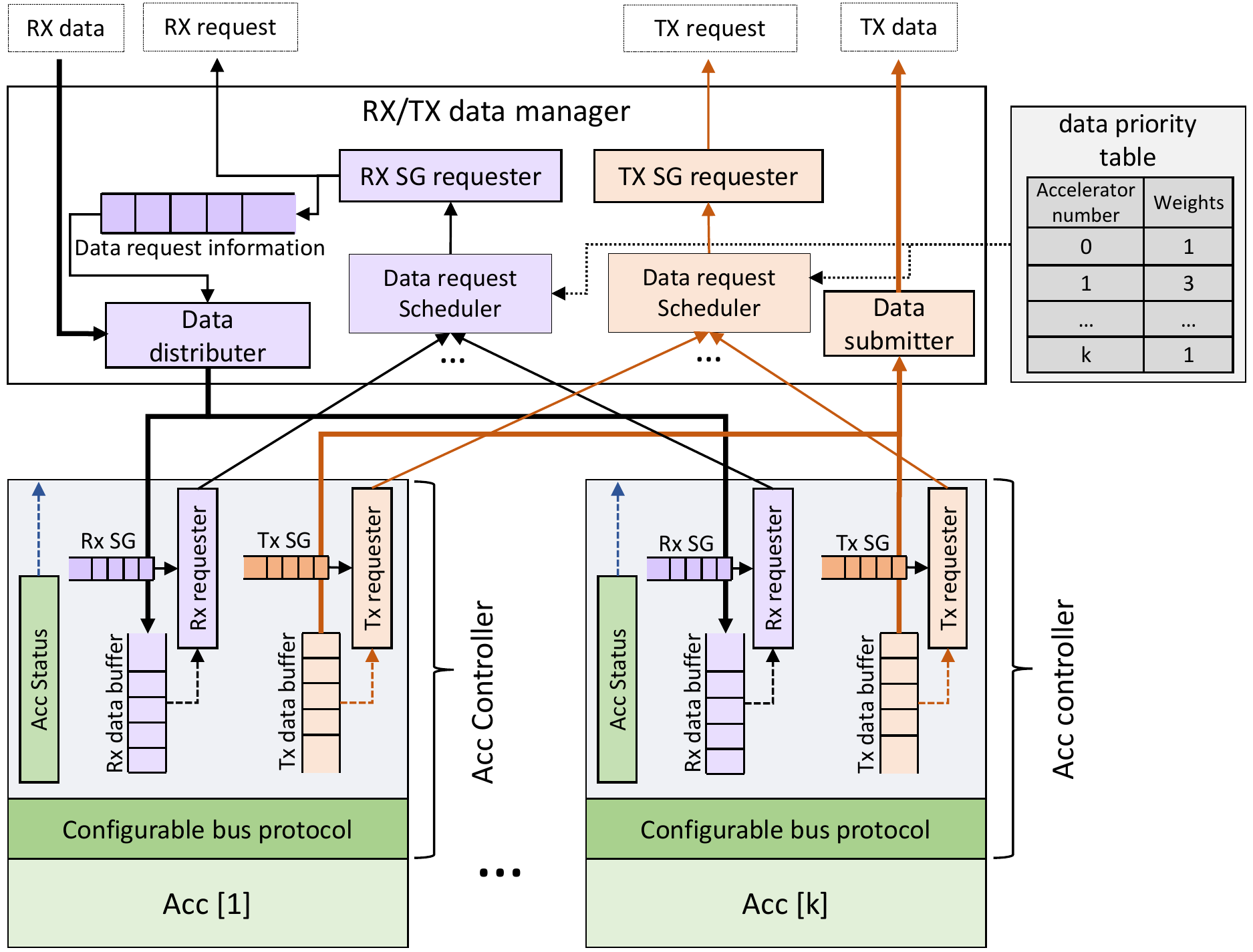}}
\caption{Accelerator controller and interleaved RX/TX scatter-gather manager}
\label{scatter_gather}
\end{figure*}

\subsection{Accelerators Controller}
\textbf{Acc Controller:}
In an accelerator controller, for each input, one RX data buffer and for each output, one TX data buffer exists. Figure \ref{scatter_gather} shows a detailed block diagram of the accelerator controller and data transfer parts.
Conventionally, RX/TX buffers must be large enough to store all the data for one accelerator request. Considering the rate of input/output data from PCIe and the rate of the data process in the accelerator, buffers could be more optimized. However, the optimization requires a careful profiling of the accelerators processing rate and the data rate transfer. On the other hand, still for most of the accelerators with a low processing rate, the size of the buffers would be relatively large. Defining large buffers in the BRAMs for the accelerator invocation framework does not leave enough space for designers to place more accelerators in an FPGA.

To overcome this problem, we define one RX scatter-gather queue and one TX scatter-gather queue for each input and output, respectively. Thus, each accelerator controller stores the whole list of scatter-gather elements. Then, an accelerator issues one RX request if there is enough space available in the RX data buffer, also issues one TX request if there is enough data available in the TX buffer. This mechanism allows defining RX/TX data buffers with much smaller sizes. This size must be at least equal to the size of one memory page of the host which is the maximum length of one scatter-gather element. To prevent an accelerator stall due to waiting for RX data or a free TX buffer, we define buffers with the size of a few numbers of memory pages.

Other than resolving the problem with large buffer sizes, our mechanism of handling scatter-gather list allows providing a scheduling strategy for serving different accelerators. It allows serving scatter-gather elements from different accelerators in any order. We provide a scheduler in the data management part (section \ref{data_transfer}) based on a configurable priority list.

\subsection{Data Transfer}\label{data_transfer}
The data transfer part is responsible for providing RX/TX data for accelerators. Together with the accelerator controller part, the RX/TX data would be fetched/submitted from/to the DMA engine for each scatter-gather element.
It is notable that the path for RX and TX data is completely separated and at the same time an RX and a TX request can be responded to and the data movement be accomplished.
In the following different components of the data transfer part are introduced.

\textbf{Data Request Scheduler}
This component uses a \textit{data priority table} to serve different accelerators. There are two different data request scheduler, one for RX data and one for TX data. Algorithm \ref{sg_scheduler} represents the functionality of the data request scheduler.
In Algorithm \ref{sg_scheduler}, $acc\_weight[acc]$ represents the priority weight related to accelerator number $acc$ and this weight comes from the data priority table. If $acc\_req[acc]$ is equal to $1$ it means that accelerator number $acc$ has a data transfer request and is waiting to be served. Accelerator number $acc$ is served when an ACK it receives an $ACK$ from the scheduler.
The \textit{data priority table} can be configured through submitting a configuration command which is provided by software developers. In the experimental section we provide the results for two different scheduling strategies: 1) fair bandwidth sharing among accelerators and 2) weighted data rate distribution among accelerators.

\begin{algorithm}
\caption{scatter-gather scheduler}
\label{sg_scheduler}
\KwIn{byte acc\_weight[k], bool acc\_req[k]}
\KwOut{bool accepted\_req[k]}
$acc \gets 0$\;
  \While{true}{
    \For{$i:0\ to\ acc\_weight[acc]$}{
        \If{$acc\_req[acc] = 1$}{
            Accept the request and send an $ACK$ to the accelerator $acc$\;
        }
    }
    $acc \gets next\ acc$\;
  }
\end{algorithm}

\textbf{RX/TX SG Requester:}
These components submit a request related to one scatter-gather element to the DMA.
The request includes an address and a length.

\textbf{Data Request Information:}
For each RX request, a data request information queue stores the information related to the request to be used when the corresponding data arrives. This information allows the \textit{data distributor} component to submit the data to the correct accelerator.

\section{Experimental Results}\label{Results}



\subsection{Experimental setup}
To synthesize and implement UltraShare, we use Xilinx\textsuperscript{\textregistered} Vivado\textsuperscript{\textregistered} 2018 design tool. We exploit a 7v3-alpha-data board which has a Xilinx\textsuperscript{\textregistered} Virtex 7 FPGA with a PCIe Gen3 connector. Our host is a PC with an Intel\textsuperscript{\textregistered} Core\textsuperscript{TM} i5-4590 CPU @ 3.30GHz.

We have implemented UltraShare in the pure Verilog programming language. Thus, UltraShare is not limited to any specific vendor tool or platform to be used. We have deployed a command-based data interface scheme in our software stack similar to \cite{mqmaiICCD18}, including multi-core parallel access to accelerators.  UltraShare can easily be used with other available command-based software platforms to access FPGA accelerators such as Xilinx SDAccel. It only requires the software stack to submit single commands that follow UltraShare command structure including an accelerator type field for managing accelerators.

\begin{figure}
\centerline{\includegraphics[scale=0.83, clip, trim=11cm 4.3cm 2cm 18cm]{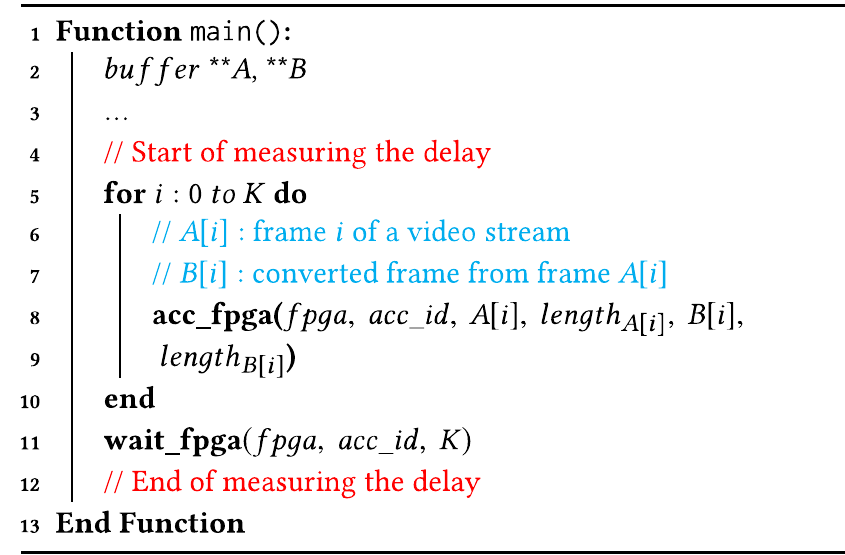}}
\caption{pseudo-code of an application invoking FPGA accelerators}
\label{pseudocode}
\end{figure}

Figure \ref{pseudocode} shows the pseudo-code of the application that we used to call the accelerators. We use two APIs for calling accelerators and waiting for their completions. We measure the throughput of the accelerators by measuring the end-to-end delay of processing requests. In Figure \ref{pseudocode}, line 4 and 12 are the places that we start and finish measuring the end-to-end delay.

\subsection{Benchmarks}
To evaluate UltraShare, we exploit two streaming accelerators: an image/video processing accelerator and a network packet encryption algorithm.

\textbf{Image processing:}
An image/video processing accelerator receives a stream of image/video frames, process them, and generates the associated stream of outputs. A wide range of image processing algorithms applies a two-dimensional stencil computation over streams of frames.
We exploit a standard image processing IP core from Xilinx\textsuperscript{\textregistered} which is an RGB to YCbCr converter with the standard AXI4-Stream interface. This accelerator can be configured, before the implementation, for different resolutions of images/videos. We define three different type of RGB to YCbCr converter, resolution $240\times180$, resolution $480\times320$, and resolution $960\times640$. While for these accelerators the computation algorithm is the same, the input/output sizes and the computation latency over a single input are different.

\textbf{Network packet encryption:}
Encryption is a streaming computation intensive algorithm that can be a good candidate to be accelerated on FPGAs. We use an AES-128 encryption algorithm from MVD\textsuperscript{\textregistered} cores over different videos with different resolutions. Unlike the RGB to YCbCr accelerator, the same AES accelerator can operate over different input sizes.

\subsection{Results}

\subsubsection{\textbf{Dynamic accelerator allocation}} To explore the impact of dynamic versus static accelerator allocation, we compare UltraShare with Riffa \cite{riffa}. Riffa is the only open-source framework which is available to be used. While ST-Accel, a recently proposed framework, has automated the process of generating and connecting accelerators to the applications, the mechanism of accelerator allocation and data transfer in ST-Accel is very similar to Riffa.

\begin{figure}
    \centerline{\includegraphics[scale=1.45]{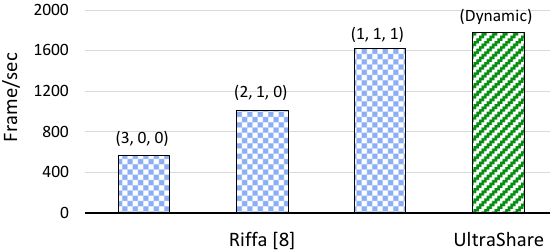}}
    \caption{Impact of dynamic vs. static accelerator allocation. UltraShare vs. Riffa}
    \label{dynamic_vs_static}
\end{figure}

Riffa is not capable of handling multiple requests from different applications to a single accelerator. Thus, to compare with Riffa, we use one multi-threaded application and use a semaphore mechanism to manage requests to the same accelerators. Figure \ref{dynamic_vs_static} shows the total throughput in processed frames per second for an application with three simultaneous threads requesting for two instances of a single type of accelerator. The chosen accelerator is an RGB to YCbCr $480\times 360$. In Figure \ref{dynamic_vs_static}, for Riffa, different scenarios of static accelerator allocation are shown on top of the related bars. For example, (3, 0, 0) means that all the three threads are requesting only for the first accelerator and (2, 1, 0) means two of the threads are requesting for the first accelerator and one of the threads is requesting for the second accelerator. Comparing to the worst case of the static accelerator allocation in Riffa, we observe more than 3x improvement in throughput. It is notable that this is just a simple scenario to show the impact of a dynamic accelerator allocation. In a more complicated scenario, a static accelerator allocation can drastically degrade the performance.

\subsubsection{\textbf{multi-queue grouping accelerators}}
To show the impact of multi-queue grouping accelerators on removing accelerators idle times, we implemented three types of accelerators: two from RGB to YCbCr converter, for resolutions $240\times180$ and $480\times320$, and one AES accelerator that we submitted video frames with the resolutions of $240\times180$ to it. From each of these accelerator types, we implemented 3 instances. Thus, totally 9 accelerators are implemented on our FPGA.

Table \ref{remove_idle} compares the throughput of UltraShare versus a non-grouping single-queue implementation. As we described in section \ref{single_vs_multi_queue} the multi-queue mechanism that is proposed by UltraShare decreases the idle times of accelerators and allows them to process a request when at least one request is available.
In this experiment, we used three different applications, each requesting to one of the accelerator types. In a single-queue non-grouping implementation, the slowest accelerators will block other accelerators to be assigned to the available requests. Thus, all the accelerators will be limited to the throughput of the slowest accelerator. It is notable that in our experiment, RGB to YCbCr $240\times180$ accelerator has a slightly higher throughput. The reason is that for this accelerator, due to smaller input sizes, the associated user application can prepare and submit more requests comparing to the other applications. Thus, more requests from this application will be ended in the shared command-queue. As Table \ref{remove_idle} shows UltraShare with a uniformed priority weights provides around 8x improvement in the throughput of the fastest accelerator type by removing the idle times caused by the slower accelerators.

\begin{figure}
\centerline{\includegraphics[scale=1.5]{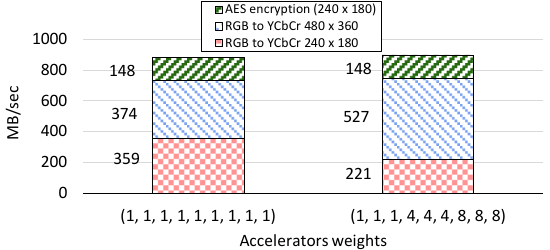}}
\caption{PCIe bandwidth sharing among the accelerators for different values of scatter-gather requests weights}
\label{PCIe_sharing}
\end{figure}

Table \ref{remove_idle} also represents a priority based PCIe bandwidth sharing provided by the \textit{data request scheduler} component. Table \ref{remove_idle} shows that how changing data priority weights, presented in Algorithm \ref{sg_scheduler}, can adjust the throughput of the accelerators when the PCIe bandwidth is the bottleneck. Figure \ref{PCIe_sharing} shows the distribution of PCIe bandwidth among the accelerators for the uniformed weights, represented as (1, 1, 1, 1, 1, 1, 1, 1, 1), and a throughput ratio-based weights, represented as (1, 1, 1, 4, 4, 4, 8, 8, 8). As Figure \ref{PCIe_sharing} shows, a uniform priority weights provides a fair bandwidth sharing among the accelerators; while, by changing the priorities we can allocate more bandwidth to specific accelerators. It is notable that the throughput of the AES accelerator is limited to its computation; thus, it cannot use its dedicated bandwidth portion; thus, the scheduler allows the other accelerators to use more bandwidth to fully utilize the PCIe bandwidth.

\begin{table}
    \caption{Throughput of different accelerators for UltraShare vs. single-queue non-grouping structure}
    \centerline{\includegraphics[scale=0.96]{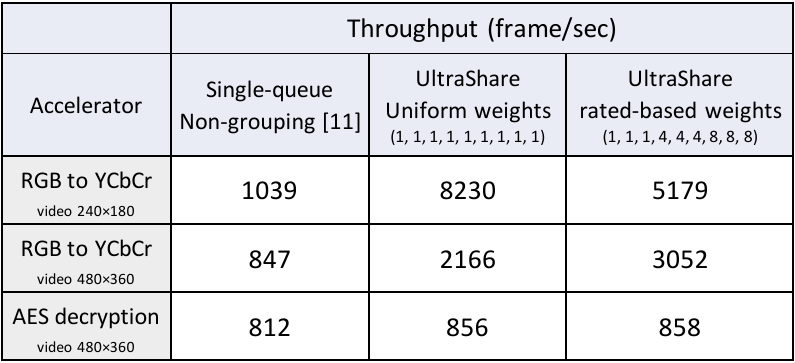}}
    \label{remove_idle}
\end{table}

\subsubsection{\textbf{Scalability}}
\begin{figure}
\centerline{\includegraphics[scale=0.7]{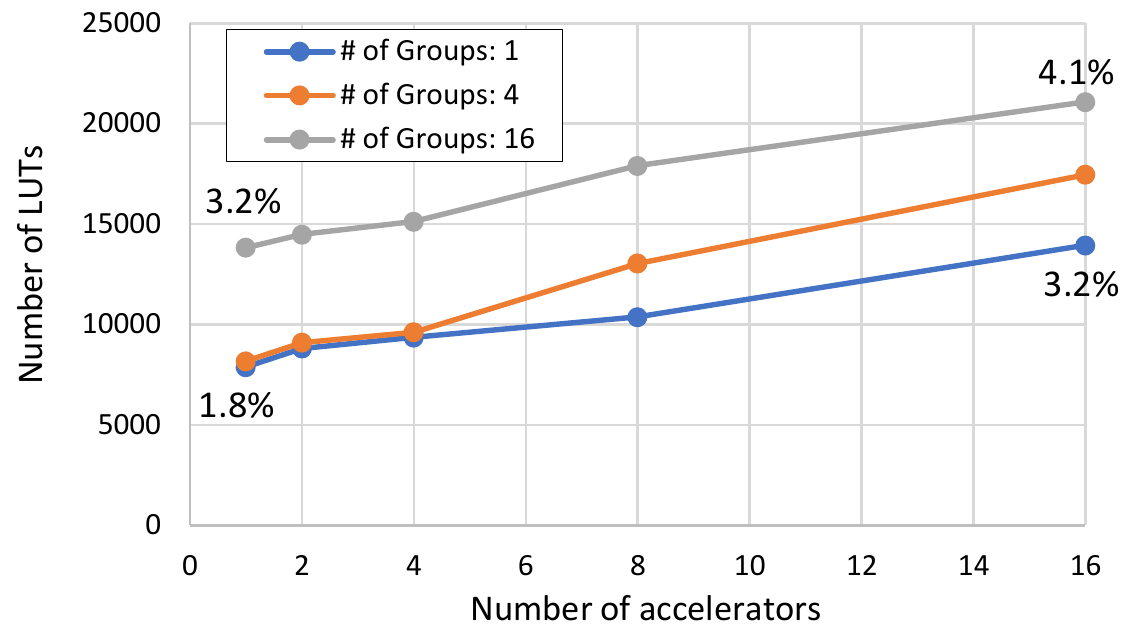}}
\caption{LUT utilization}
\label{LUT}
\end{figure}

\begin{figure}
\centerline{\includegraphics[scale=0.7]{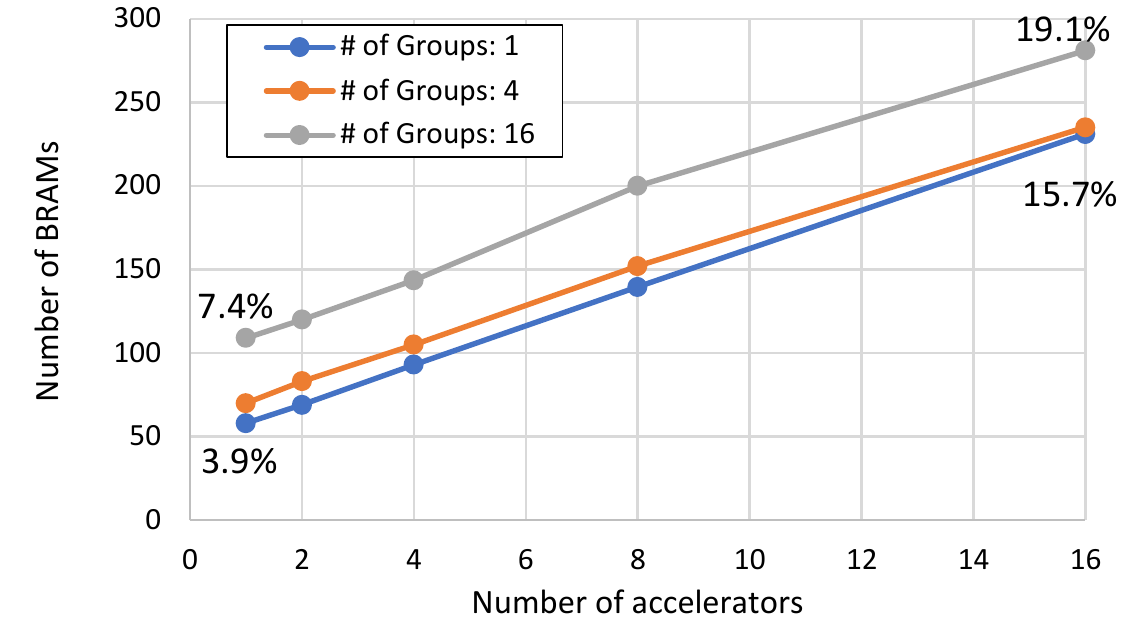}}
\caption{BRAM utilization}
\label{BRAM}
\end{figure}

To show the scalability of UltraShare, we measured the resource utilization with a various number of accelerators and a various number of accelerator groups. Among all the resources, only the utilization and variation of LUTs and BRAMs are considerable. Figure \ref{LUT} and Figure \ref{BRAM} show the number of utilized LUTs and BRAMs, respectively (the given percentages are based on our Xilinx\textsuperscript{\textregistered} Virtex 7 FPGA). As Figure \ref{LUT} shows increasing the number of accelerators and accelerator groups have almost the same effect on the number of LUTs and it is linear with a low slope.
The variation of BRAM utilization is higher because all the buffers and queues are implemented in the BRAMs. As Figure \ref{BRAM} shows, the number of accelerators has a greater impact on the BRAM utilization and it is because of all the buffers and queues that are used in each accelerator controller versus only the number of command queue utilization of BRAMs for each additional accelerator group.

\subsubsection{\textbf{Exploiting dynamic parallelism}}
\begin{figure}
\centerline{\includegraphics[scale=0.66]{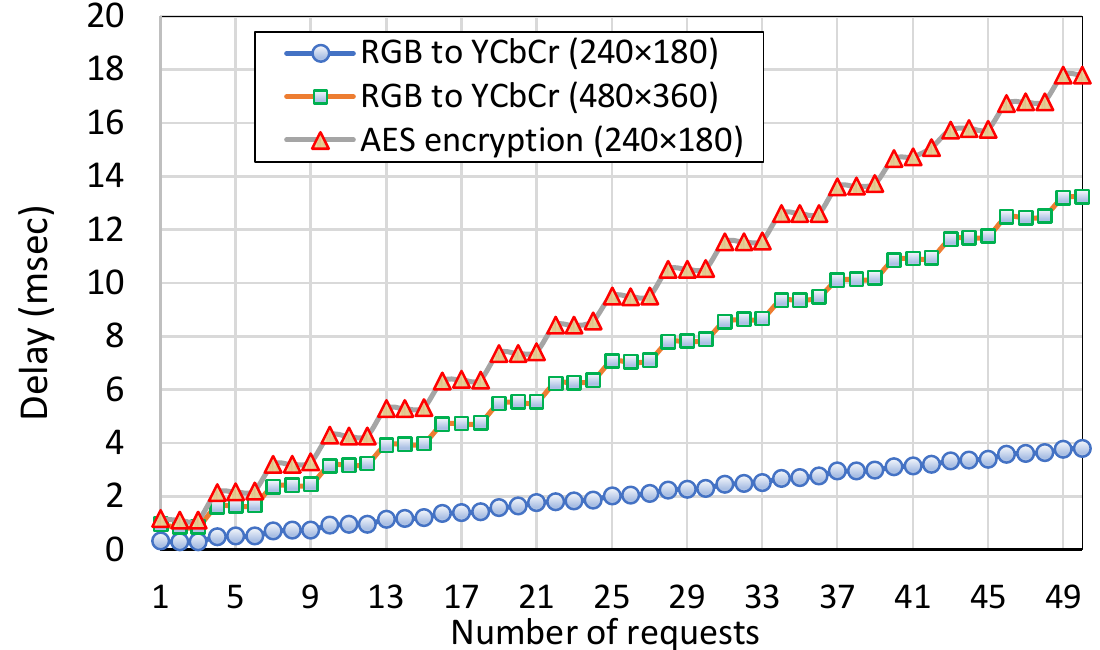}}
\caption{Exploiting parallelism in UltraShare}
\label{result_parallelism}
\end{figure}

To show the capability of supporting parallelism that UltraShare supports, we provided three separate experiments. In each experiment, we implemented three instances of only one type of accelerators. 
Figure \ref{result_parallelism} shows the end-to-end delay of submitting requests from a single application to the target accelerators.
As Figure \ref{result_parallelism} shows, when the number of requests increases, requests will be distributed among the accelerators and for each factor of three, there is a jump in the end-to-end delay. The reason is that there are no more accelerators available for a new request after all the three accelerators are processing three different requests. It means that the forth request needs to wait until at least one accelerator is idle. Thus, the delay is increased after passing a factor of three request which is the number of accelerators.

\subsubsection{\textbf{Accelerator sharing}}
\begin{figure}
\centerline{\includegraphics[scale=1.4]{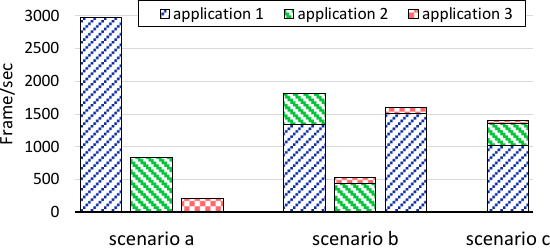}}
\caption{AES accelerator sharing among different applications submitting three different video resolutions}
\label{acc_sharing}
\end{figure}

\begin{figure}
\centerline{\includegraphics[scale=1.4]{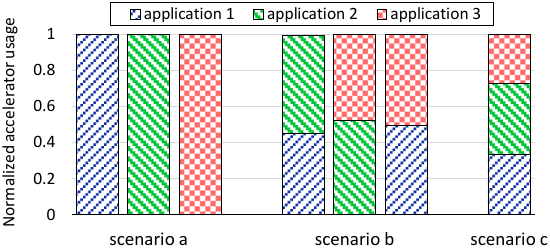}}
\caption{Normalized AES accelerators usage by three different applications submitting three different video resolutions}
\label{acc_usage}
\end{figure}

To show the accelerator sharing feature of UltraShare among different applications, we implemented three instances of AES encryption accelerator on the FPGA. Then we provided three different applications: \textit{application 1} that submits requests for a video with the resolution of $240\times180$, \textit{application 2} that submits requests for a video with the resolution of $480\times360$, and \textit{application 3} that submit requests for a video with the resolution of $9600\times640$. Then we considered three different scenarios (Figure \ref{acc_sharing}). In \textit{scenario a}, we ran only one of the applications at a time and measured the throughput of the accelerators for each application. In \textit{scenario b}, we ran two different applications simultaneously and we considered all the three possible combinations of two applications from three applications. In \textit{scenario c}, we ran all the applications simultaneously. Considering the throughput of the processed frames for each application in these three scenarios, presented in Figure \ref{acc_sharing}, we can see how the accelerators are evenly shared among the applications. Although due to the different input sizes the number of processed frames for the different applications are different (when the input size is larger, it takes longer time to be processed), Figure \ref{acc_usage} \textit{scenario c} shows that the accelerator usage for all the three applications are equal. On the other word, the difference in throughput is due to the different request sizes which require different computation latency.
\section{Conclusions}\label{Conclusions}
In this paper, we proposed UltraShare, an FPGA-based accelerator hardware controller to enable dynamic accelerator sharing among multiple host applications.
UltraShare provides a scalable dynamic accelerator allocation scheme to exploit dynamic parallelism for the requests from a single application. Using an \textit{accelerator grouping} scheme, UltraShare removes accelerator idle times to improve the total throughput gained from a multi-accelerator system. UltraShare also deploys a single command-based request mechanism that addresses a non-blocking accelerator sharing environment for different host appellations to share FPGA accelerators.
UltraShare follows a scatter-gather based point-to-point data transfer from the main memory to/from the FPGA to avoid data-intensive streaming accelerators being stalled due to an inefficient data transfer in a hierarchical memory model.
Experimental results show that in a simple scenario with 9 accelerators from 3 different types, UltraShare provides up to 8x throughput improvement for streaming applications comparing to a single-queue non-blocking implementation.

\bibliographystyle{ieeetr}
\bibliography{sample-base}

\end{document}